\documentclass[aps,twocolumn,floatfix,superscriptaddress]{revtex4-1}

\usepackage{graphicx}
\usepackage{indentfirst}
\usepackage{latexsym}
\usepackage{multirow}
\usepackage{tabls}
\usepackage{epsfig}
\usepackage{multirow}
\usepackage{subfigure}
\usepackage{comment}
\usepackage{hyperref}

\usepackage{color}
\usepackage[usenames,dvipsnames]{xcolor}
\usepackage{amssymb}

\usepackage{amsfonts}
\usepackage{amsmath}
\usepackage{bm}
\usepackage{mathrsfs}

\usepackage{algorithm}
\usepackage{algpseudocode} 
\usepackage{setspace} 

\def\d{ \delta }
\def\e{ \epsilon }

\def\cM{ {\cal M} }
\def\cN{ {\cal N} }

\def\cW{ {\cal W} }

\newcommand{\be}{\begin{equation}}
\newcommand{\ee}{\end{equation}}

\def\lsim{\mathrel{\rlap{\lower3pt\hbox{\hskip1pt$\sim$}}
    \raise1pt\hbox{$<$}}}                
\def\gsim{\mathrel{\rlap{\lower3pt\hbox{\hskip1pt$\sim$}}
    \raise1pt\hbox{$>$}}}         

\def\coordeq{ \, \mathrel{ \rlap{\hbox{\hskip-2.5pt$=$} }
    \raise4pt\hbox{$\cdot$}} \, }                

\begin{document}

\title{A sparse representation of gravitational waves from precessing compact binaries}

\def\addUMD{Center for Scientific Computation and Mathematical Modeling, and Joint Space Sciences Institute,  Maryland Center for Fundamental Physics, Department of Physics, University of Maryland, College Park, MD 20742, USA}

\def\addCaltech{Theoretical Astrophysics, California Institute of Technology, Pasadena, CA, 91125, USA}

\author{Jonathan Blackman}
\affiliation{\addCaltech}

\author{Bela Szilagyi}
\affiliation{\addCaltech}

\author{Chad R.\,Galley}
\affiliation{\addCaltech}

\author{Manuel Tiglio}
\affiliation{\addUMD}
\affiliation{\addCaltech}

\begin{abstract}
	Many relevant applications in gravitational wave physics 
share a significant common problem: the seven-dimensional parameter space of gravitational waveforms from precessing compact binary inspirals and coalescences is large enough to prohibit covering the space of waveforms with sufficient density.
We find that by 
using the reduced basis method together with a parametrization of waveforms based on their phase and precession,  we can construct ultra-compact yet high-accuracy representations of this large space. As a demonstration, we show that less than $100$ judiciously chosen precessing inspiral waveforms are needed for $200$ cycles, mass ratios from $1$ to $10$ and spin magnitudes $\le 0.9$. In fact, using only the first $10$ reduced basis waveforms yields a maximum mismatch of $0.016$ over the whole range of considered parameters. 
We test whether the parameters selected from the inspiral regime result in an accurate reduced basis when including merger and ringdown; we find that this is indeed the case in the context of a non-precessing effective-one-body model.
This evidence suggests that as few as $\sim 100$ numerical simulations of binary black hole coalescences may accurately represent the seven-dimensional parameter space of precession waveforms for the considered ranges.
\end{abstract}

\maketitle

Gravitational radiation produced by 
stellar-mass compact binaries of neutron stars and/or black holes
 are expected to be the main signals detected by the advanced generation of gravitational wave detectors \cite{Harry:2010zz,Losurdo:2008zz,Somiya:2011np,Indigo_web}. Detecting these signals and estimating the parameters of their sources require the ability to sufficiently sample the space of precessing compact binary waveforms. A compact binary intrinsically depends on its mass ratio and the spin angular momentum components of each body, which forms a $7$-dimensional space for gravitational waveforms \footnote{Gravitational wave detectors have a finite frequency bandwidth that introduces a total mass scale, thus adding an $8^{\rm th}$ parameter that we will ignore. We also focus on quasi-circular inspirals.}. 
 
Much progress has been made in sampling the subspace of non-spinning compact binary waveforms over the last decade. However, many relevant applications, from data analysis for gravitational wave searches and parameter estimation studies to numerical relativity simulations of binary black hole coalescences, face a common challenge. In particular, going from the non-spinning subspace to the full $7$d space naively requires prohibitively more samples for the same coverage simply because the volume of the space grows exponentially with dimension. As a result, the general consensus in the gravitational wave community is that the computational complexity 
associated with building template banks for matched-filter searches, with making parameter estimation studies, and with modeling 
precessing compact binaries by expensive numerical simulations is intractable (e.g., see \cite{Hannam:2013pra}). 
This phenomenon entails what is called the {\it curse of dimensionality} \cite{Bellman:1961}. 

In this paper, we show how to beat the curse of dimensionality for precessing compact binary inspirals.
We find that only $50$ judiciously chosen waveforms are needed to represent the entire $7d$ space with an accuracy better than $10^{-7}$ for $200$ cycles, mass ratios $q \in [1,10]$, dimensionless spin magnitudes $\|\vec{\chi}_{1,2} \| \le 0.9$, and through $\ell = 8$ spherical harmonic modes.
Using only the first $10$ of these select waveforms yields a maximum representation 
error $\lesssim 1\%$.
The results of this paper suggest that for any given parameter range a remarkably small number of numerical relativity simulations of precessing binary black holes, if judiciously chosen, are sufficient to {\it accurately represent any other precession waveform in that range}. We expect these results to be useful also for gravitational wave matched-filter searches and parameter estimation studies for compact binary coalescences.

~\\
\noindent{\it Beating the curse of dimensionality}.--
Previous studies have shown that non-precession subspaces of the full $7$d waveforms space $\cW$ can be represented by linear spaces spanned by a
 relatively compact set of inspiral \cite{Field:2011mf, Field:2012if}, ringdown \cite{Caudill:2011kv}, and  inspiral-merger-ringdown (IMR) \cite{Field:2013cfa} waveforms, which form a {\it reduced basis} (RB). 
The reduced basis waveforms are found by training a {\it greedy algorithm} \cite{Binev10convergencerates,Devore2012} to learn from a given discretization of $\cW$ which are the most relevant waveforms for representing elements of $\cW$ with regard to a given error measure. See \cite{Field:2011mf} for more details.
The number of RB waveforms for non-precessing inspirals hardly grows from two to four parameter dimensions thereby explicitly beating the curse of dimensionality \cite{Field:2012if}. Of further interest is that {\it precession} subspaces of $\cW$ carry significant redundancy and are amenable to dimensional reduction as found in \cite{Galley:2010rc}.
Those results strongly suggest that one may beat the curse of dimensionality in the full $7$d waveform space. 

In this paper, we outline how to construct a very compact but highly accurate RB of precession waveforms. 
We consider the following specifications on the $7$d waveform space:
\be
	q\in[1,10] ~ , ~~ \| \vec{\chi}_{1,2}\| \in [0,0.9] ~,  ~~ 200 \text{ cycles}  .
 	\label{eq:specs} 
\ee
where $q = m_1 / m_2 \ge 1$.
These were chosen based partially on practical limitations of binary black hole simulations.
 However, the general message of this paper does not depend on our choice.

~\\ 
\noindent{\it Key ingredients}.--Our construction of a very compact (or sparse) reduced basis representation of precession waveforms depends on several key steps \footnote{The last two ingredients can be viewed as aspects of nonlinear dimensional reduction and manifold learning, which aim to reveal the intrinsic dimensionality of large amounts of data (e.g., see \cite{NLDRLeeVerleysen}). 
}: (1) A randomized resampling strategy \cite{EPFL-ARTICLE-190659} for training the greedy algorithm on the $7$d waveform space; (2) A frame that rotates with the binary's precession; and (3) A physically motivated parametrization of precession waveforms.

The first key ingredient is a modification of the standard greedy algorithm \cite{Field:2011mf}.
 In its simplest inception, the greedy algorithm learns which waveforms can linearly span the space of interest 
 in a nearly optimal way \cite{Binev10convergencerates,Devore2012}, starting from a sufficiently dense set of waveforms called a training set or space.
 However, the curse of dimensionality prevents us from sampling the waveform space with sufficient coverage to build a reliable training set. To overcome this, we randomly resample the $7$d space from a uniform distribution using a fixed number $K$ of waveforms at each iteration of the greedy algorithm. These waveforms constitute the training set at the current iteration. 
Because the $7$d space is resampled at each iteration by different waveforms, the maximum error from projecting waveforms onto the current basis at the $j^{\rm th}$ step is actually measuring this error over an effective training set with $j \times K$ randomly distributed waveforms. As more iterations are made, more of the $7$d space is sampled and the more accurate the RB becomes.
This is a simple implementation of more powerful techniques introduced in Ref.~\cite{EPFL-ARTICLE-190659}. 

For our studies, we randomly and uniformly resampled $K \le 36,\!000$ waveforms at each iteration of the greedy algorithm. 
We began our studies with small $K$ and increased each sample size up to $K=36,\!000$, for which the total number of RB waveforms was robust and independent of $K$. The largest training set used in our studies included
 more than $3\times 10^6$ randomly selected waveforms.

The second key ingredient is to work in the binary's precessing frame instead of the usual inertial one. Specifically, we generate post-Newtonian (PN) precession waveforms in the time-domain using the {\it minimally rotating} frame of Refs.~\cite{Boyle:2011gg, Boyle:2013nka}. 
In this frame, a precession waveform appears qualitatively similar to waveforms from non-spinning binaries in their inertial frame \cite{Schmidt:2012rh, Lundgren:2013jla, Pekowsky:2013ska, Boyle:2013nka, Hannam:2013oca, Hannam:2013pra}. In the minimally rotating frame, waveforms have a weaker dependence on parameters than they do in the inertial frame. 
The rotation involved in going from the minimally rotating frame to the inertial one and vice versa can be accounted for by any convenient representation of the $SO(3)$ group. 

The third key ingredient, and perhaps the most crucial, is that we choose to parametrize precession waveforms by their phase instead of by time or frequency.
To motivate this choice we momentarily consider
 the frequency-domain waveform (in the stationary phase approximation) for a non-spinning binary inspiral at leading order (``$0$PN'') in the PN approximation,
\be
	h( f; \cM) = A \cM^{5/6} f^{-7/6} e^{i \Phi_0 (f; \cM)} , \label{eq:0PN}
\ee
where $\cM = M \nu^{3/5}$ is the chirp mass, $M$ is the total mass, $\nu$ is the symmetric mass ratio, $A$ is a constant independent of the binary's intrinsic parameters, and 
\be
	\Phi_0 (f; \cM) \equiv \frac{3}{128} \big( \pi \cM f \big)^{-5/3} . \label{eq:0PNphase}
\ee
Reparametrizing (\ref{eq:0PN}) by its phase, now taken as the independent variable, gives 
\be
	H (\varphi; \cM) \equiv h(F(\varphi); \cM) =  A' \cM^2 \varphi^{7/10} e^{i \varphi} \,    \label{eq:0PN_phase}
\ee
with $A' = A \pi^{7/6} (128/3)^{7/10}$ and $F(\varphi)$ from solving $\Phi_0(f$=$F) = \varphi$.  In this phase-domain, all  waveforms are proportional to each other, thus constituting a $1$d space. 
In fact, performing the greedy algorithm analytically (versus numerically) returns a single basis waveform that {\em exactly} represents all such waveforms in the continuum.
This is the intrinsic dimensionality of the problem as has long been known because $0$PN waveforms only depend on the chirp mass. Therefore, a single reduced basis waveform spans the whole $0$PN waveform space. To close the system, we also need to represent the mapping between the phase and frequency domains, 
\begin{align}
	F(\varphi; \cM) = \frac{ 1}{ \pi \cM } \left( \frac{128 \varphi}{3}  \right)^{-3/5}, 
\label{eq:inverse1}
\end{align}
 using a separate basis. 
As we see again, the frequencies for different chirp masses are all proportional to each other. Therefore, 
any $0$PN waveform, as a function of frequency, is represented by {\em one} reduced basis waveform
through the non-linear transformation in (\ref{eq:0PNphase}). 

For the sake of comparison,
we implemented a standard greedy algorithm following \cite{Field:2011mf} using $0$PN waveforms parametrized by frequency (not phase)
for binaries with a fixed total mass and with mass ratios and number of cycles as listed in (\ref{eq:specs}). We found that $152$ RB waveforms are required to reach numerical round-off errors in representing any waveform in this $1$d space. Even to reach an error of about $1\%$ requires $138$ RB waveforms. Therefore, using the phase parametrization results in a single RB waveform for {\it exact} representation whereas a frequency parametrization can yield a much larger RB for {\it approximate} representation. 

Part of the reason why using waveforms in the phase domain (or $\varphi$-domain) is advantageous is because the oscillations in two waveforms {\it always} cancel in the scalar product used to measure the projection error onto the RB in the greedy algorithm,
\begin{align}
	\big\langle H_{\cM_1}, H_{\cM_2} \big\rangle_{\varphi} \equiv \int_{\varphi_{\rm min}}^{\varphi_{\rm max}} {\hskip-0.15in} d\varphi \, H(\varphi; \cM_1) H^*(\varphi; \cM_2) .
	\label{eq:0pndp0} 
\end{align}
For $0$PN waveforms this results in a very smooth dependence on the chirp masses since (\ref{eq:0pndp0}) is $\propto \cM_1^2 \cM_2^2$. Similarly, the waveform frequency as a function of phase (\ref{eq:inverse1}) has a very smooth dependence on them as well. 

Higher PN orders include more physics, such as the nonlinearity of general relativity and spin-orbit, spin$1$-spin$2$, and self-spin interactions, that depend on all $7$ intrinsic parameters. These contributions add more structure to the waveforms but only weakly depend on the parameters. This is especially true in the $\varphi$-domain and, as discussed below, 
we also find this holds through the merger and ringdown phases where the PN expansion parameter is no longer small. Consequently, the inverse function $F(\varphi)$ (or $T(\varphi)$ if in the time domain) retains the weak dependence on intrinsic parameters. As there is thus only ever a weak parameter dependence, one may expect %
to find only a relatively small number $\cN$ of RB waveforms, possibly as few as $\cN = O(d)$. 

~\\
\noindent{\it Method outline}.--In this paper, we use 3.5PN precessing inspiral waveforms. We solve the PN equations (see Ref.~\cite{Bohe:2012mr} and references therein)
using the approach of Refs.~\cite{Boyle:2011gg,Boyle:2013nka} where the waveforms themselves are solved in a frame that minimizes the binary's precession, along with a rotation operator represented by unit quaternions to track this frame relative to the fiducial inertial frame. 
All waveforms in this minimally rotating frame are normalized to unity, and the initial orbital phases are aligned.
It is convenient to decompose the waveform into spin-weighted spherical harmonic modes \cite{Thorne:1980ru} characterized by $(\ell, m)$. We use the phase associated with the $(\ell, m) = (2,2)$ mode to parametrize the waveform but other choices are possible. A precession waveform $h (t)$ in the inertial frame is thus decomposed in the following way,
\begin{align}
	h (t) \rightarrow \big( \{ H_{\ell m} (\varphi) \}, T(\varphi), Q(\varphi) \big)
\label{eq:decomp1}
\end{align}
where $H_{\ell m}$ is a spin-weighted spherical harmonic mode in the minimally rotating frame, $T(\varphi)$ is the function relating the $(2,2)$ phase to the time coordinate, and $Q$ is the unit quaternion describing the rotation back to the inertial frame.
We take into account all modes up to $\ell = 8$ and cut all waveforms off at a dimensionless frequency of $0.2$ in the $(2,2)$ modes. Finally, all waveforms contain $200$ wave cycles. 

We build a RB for each component in the decomposition (\ref{eq:decomp1}).
It is natural to use the scalar product in (\ref{eq:0pndp0}) for the $T$ and $Q$ functions but to integrate the minimally rotating waveform over the $2$-sphere so that, upon using the orthogonality of the spin-weighted spherical harmonics, 
\begin{align}
	\big\langle H_{\lambda_1}, H_{\lambda_2} \big\rangle_{\varphi} & \equiv \sum_{\ell,m}\int_{\varphi_{\rm min}}^{\varphi_{\rm max}} {\hskip-0.15in} d\varphi \, H_{\ell m}(\varphi; \lambda_1) H_{\ell m}^*(\varphi; \lambda_2) \, ,
	\label{eq:errors_precession}
\end{align}
where $\lambda_{i}$ is a tuple of parameter values.
Executing a greedy algorithm on each component in (\ref{eq:decomp1}) would result in a selection of parameter values that are different for each element.
In order to choose the same parameters for all three reduced bases, we define a total projection error $\e_{\varphi}$ through,
\be
	\e_{\varphi} (\lambda) \equiv 8\times 10^{-6} \| \d T_\lambda \|_{\varphi } ^2 + 0.5 \| \d H_\lambda \|_{\varphi }^2 + 0.0031 \| \d Q_\lambda \|_{\varphi } ^2 
\label{eq:norm_phase}
\ee
so as to receive approximately equal contributions from each component. Here, $\lambda = (q, \vec{\chi}_1, \vec{\chi}_2)$ is a tuple of $7$d parameter values,  $\delta X_\lambda = X_\lambda - P_X[X_\lambda]$ with $X$ one of the elements in (\ref{eq:decomp1}), and $P_X$ is the projection operator onto the basis for $X$. The numerical coefficients are fixed to give approximately equal contributions to the mismatch 
in the time domain and inertial frame in the case of small random perturbations. 
Binaries with periods near $200M$ lead to a small coefficient for the time function.

\begin{figure}
\includegraphics[width=\columnwidth]{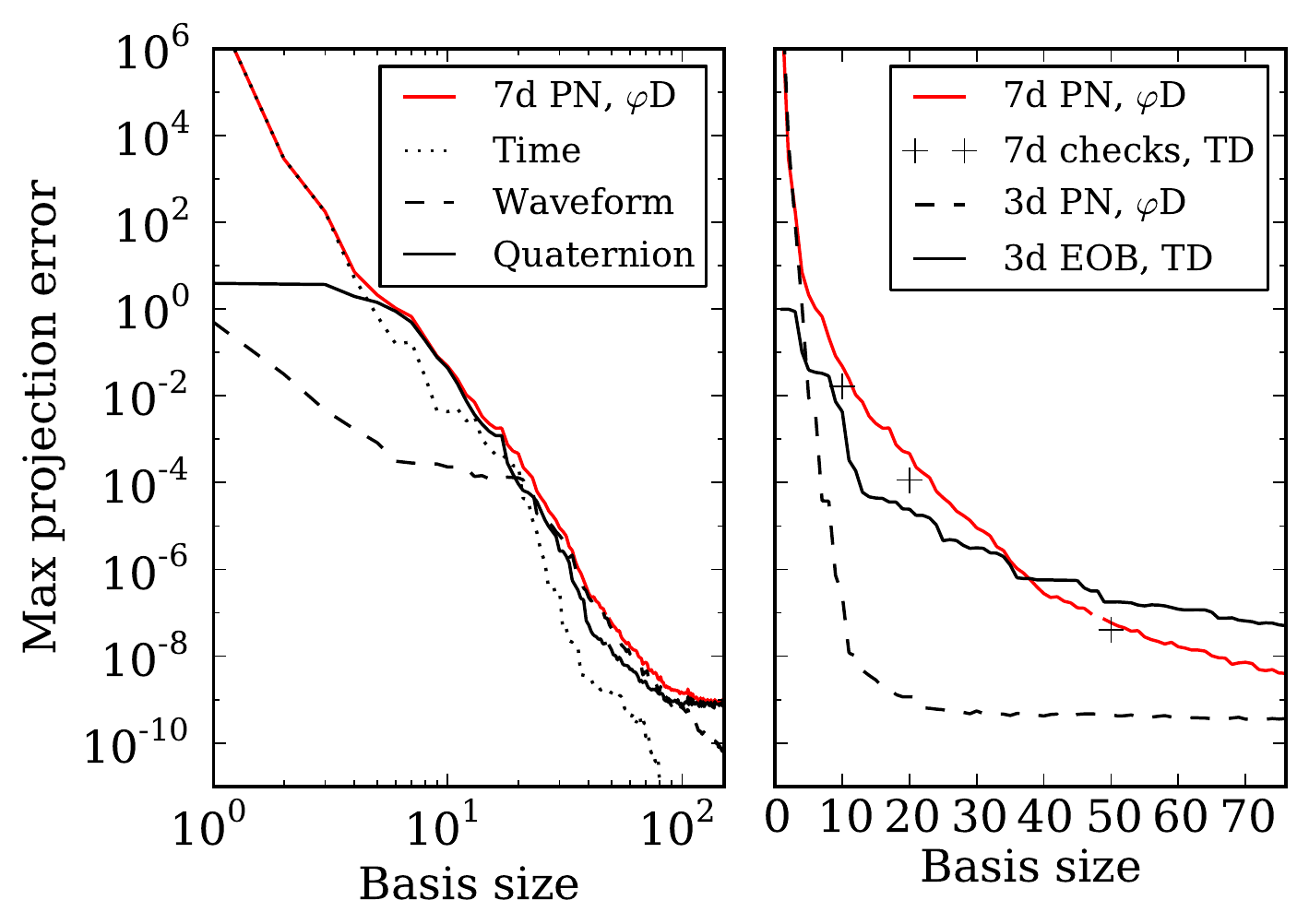} 
\caption{
{\bf Left}: Maximum $\varphi$-domain ($\varphi$D) projection error (red) from (\ref{eq:norm_phase}) for $7$d post-Newtonian precession waveforms versus basis size. The contributions from the time function (dotted), waveform in the minimally rotating frame (dashed), and quaternion (solid) are also shown. 
{\bf Right}: Maximum time-domain, inertial frame mismatches from (\ref{eq:mismatch}) for $10^7$ randomly selected waveforms ($+$) using the first $10$, $20$, and $50$ reduced basis waveforms. Also plotted are $\varphi$-domain projection errors for non-precessing PN waveforms (dashed) and the time-domain (TD) projection errors from using the latter parameter values selected by the greedy algorithm to represent EOB waveforms (solid black), which additionally include merger and ringdown phases. 
}
\label{fig:errors}
\end{figure}

~\\
\noindent{\it Results for precessing binary inspirals}.--We implemented a greedy algorithm using the three key ingredients discussed above to find RB representations for the space of precession waveforms for the ranges given in (\ref{eq:specs}). The left panel of Fig.~\ref{fig:errors} shows the maximum of the total projection error (\ref{eq:norm_phase}) found at each iteration of the greedy algorithm. We observe a power-law decay with exponent  $\approx -8$.
 The total error is not monotonically decreasing because of the constant resampling at each iteration. We observe that the maximum normalized projection error over the training set is $10^{-2}$ using $10$ basis waveforms and $\approx 4\times 10^{-8}$ for $50$.
Also shown are the contributions to the total error from the projections onto the basis of each component in (\ref{eq:decomp1}).

To measure the error in the time-domain inertial frame between a waveform $h$ and its RB approximation $h_{\rm app}$ we use the standard {\it mismatch} 
\begin{align}
	{\rm Mismatch} = 1 - {\rm Re} \,  \langle h, h_{\rm app} \rangle_t  \, ,
\label{eq:mismatch}
\end{align}
where for two functions $A, B$ the time-domain complex scalar product is $\langle A, B \rangle_t \equiv \int_{t_{\rm min}}^{t_{\rm max}} dt \, A(t) B^*(t)$.
In order to measure the quality of the RB approximation itself, we do not extremize the mismatch with respect to the relative phase and time shift between $h$ and $h_{\rm app}$. 

\begin{figure}[h]
	\includegraphics[width=\columnwidth]{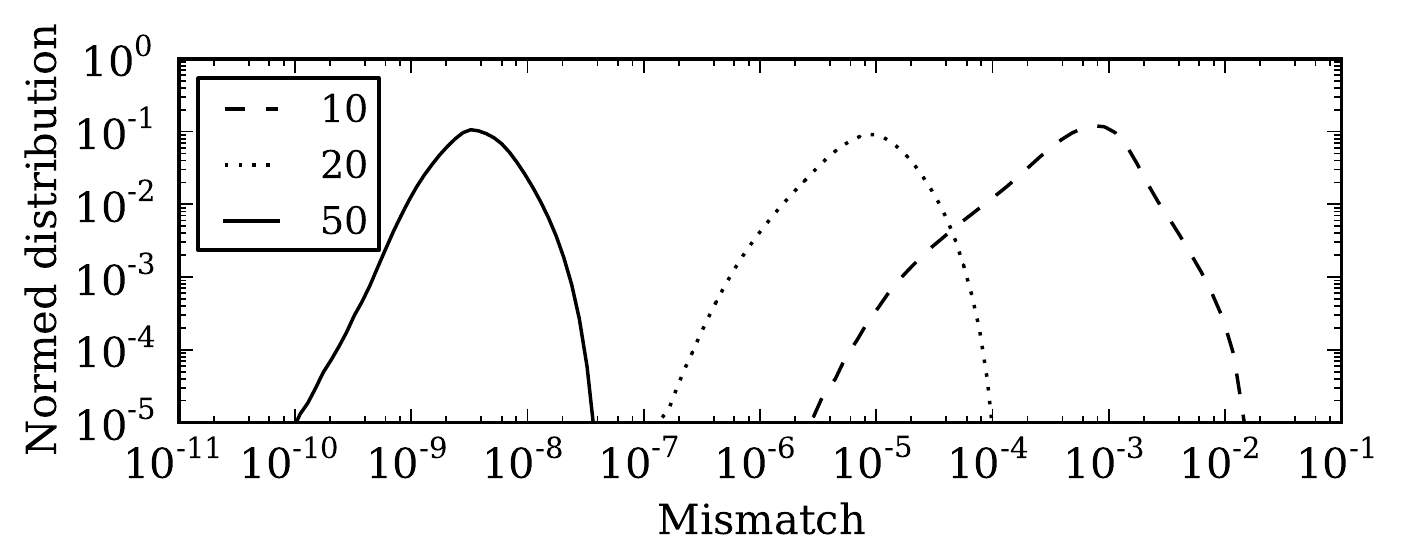}
	\caption{Distribution of mismatches from (\ref{eq:mismatch}) for more than $10^7$ randomly selected waveforms in the time-domain and inertial frame using the first $10$ (dashed), $20$ (dotted), and $50$ (solid) reduced basis waveforms. Distributions are normalized by their total samples. 
}
	\label{fig:reconstruction}
\end{figure}

It is not obvious that the basis generated using (\ref{eq:norm_phase}) from the minimally rotating frame and $\varphi$-domain will be accurate for inertial frame waveforms expressed in the time-domain. Nevertheless, we find that the $\varphi$-domain, precessing basis is highly accurate for representing time-domain, inertial-frame waveforms. The right panel in Fig.~\ref{fig:errors} shows the mismatch ($+$) from using the first $10$, $20$, and $50$ basis functions to represent more than $10^7$ randomly chosen waveforms for the same specifications in (\ref{eq:specs}). Figure \ref{fig:reconstruction} shows the distribution count of waveforms with a given error using the first $10$, $20$, and $50$ RB functions. The latter distribution has median 
$3.5 \times 10^{-9}$, mean value $4.2\times 10^{-9}$, and a maximum representation error of $4.1 \times 10^{-8}$. Using the first 10 RB functions, the maximum mismatch is $0.016$ over more than $10^7$ randomly selected waveforms.

Table \ref{tab:lowerDimensions} shows that we beat the curse of dimensionality since, for a given error, the number of RB waveforms needed to accurately represent the subspace of $\cW$ with the indicated dimension $d$ grows approximately linearly with $d$, not exponentially. 

\begin{table}[h]
\centering
\begin{tabular}{| c || c | c | c | c |}
\hline
Error & \multicolumn{4}{c|}{Basis size} \\ 
\cline{2-5}
 & $~1$d$~$ & $~2$d$~$ & $~3$d$~$ & $~7$d$~$ \\ \hline
 $~\lesssim 10^{-2}~$ & $4$ & $6$ & $7$ & $13$ \\ \hline
$~\lesssim 10^{-4}~$ & $4$ & $7$ & $8$ & $20$ \\ \hline
$~\lesssim 3 \times 10^{-8}~$ & $6$ & $15$ & $23$ & $50$ \\ \hline
\end{tabular}
\caption{The number of basis waveforms required for a desired maximum mismatch scales approximately {\it linearly} with the dimension thus beating the curse of dimensionality.  The first three dimensions considered are from mass ratio $q$ and $z$-components of the spin vectors $\vec{\chi}_{1,2}$ with $1{\rm d} \rightarrow (q)$, $2{\rm d} \rightarrow (q, \chi_{1z})$, and $3{\rm d} \rightarrow (q, \chi_{1z}, \chi_{2z})$.
}
\label{tab:lowerDimensions}
\end{table}

\begin{figure}
	\includegraphics[width=\columnwidth]{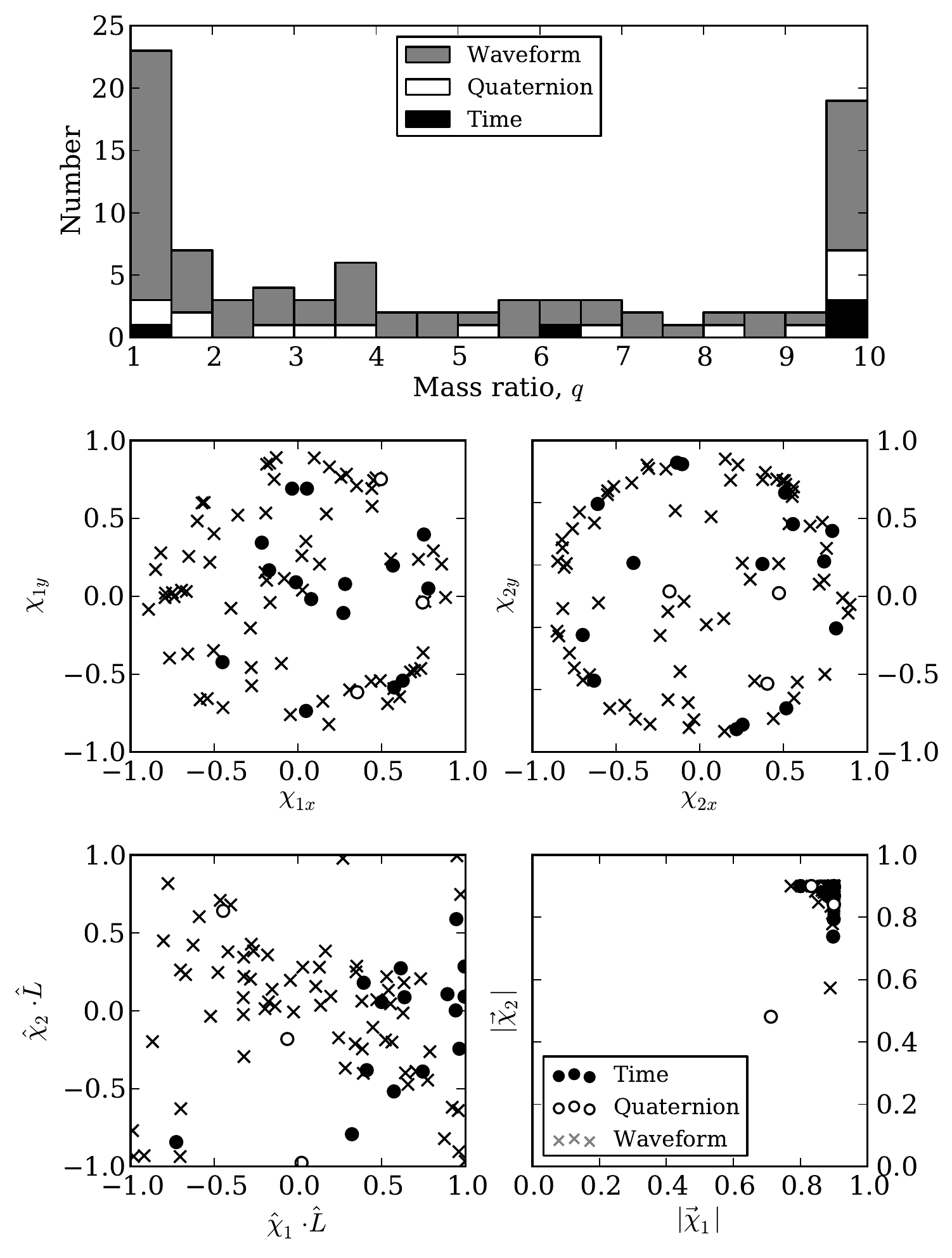} 
	\caption{Distribution of mass ratios (top), $x$-$y$ components of both spins (second row), projection of spins onto initial orbital angular momentum unit vector (bottom left), and both spin magnitudes (bottom right) as selected by our greedy algorithm. The spins' components are given in the inertial frame at the initial time.
	}
	\label{fig:q_and_spins}
\end{figure}

Figure \ref{fig:q_and_spins} shows the first 90 parameters selected by our greedy algorithm and presented according to which component -- time, minimally rotating waveform, quaternion -- is the dominant contribution to the total representation error in the left panel of Fig.~\ref{fig:errors}. The spins' components are taken at the initial time where the inertial and minimally rotating frames are equal. The mass ratios are heavily weighted towards the endpoints of the considered interval in (\ref{eq:specs}). Both spins' magnitudes tend to be in $[0.8, 0.9]$. The projections of the spins onto the initial orbital angular momentum seem to be {\it anti}-correlated, at least when the waveform contribution to (\ref{eq:norm_phase}) is dominant. We also see that the $x$-$y$ components of the spins tend to lie on a circle for the smaller mass $m_2$ while there is less clear structure for the larger mass $m_1$.

~\\
\noindent{\it From inspiral to coalescence}.--
Next, we test whether the parameters selected from the inspiral regime result in an accurate reduced basis when including merger and ringdown.
This issue has immediate relevance for building a RB for expensive numerical relativity simulations of precessing binary black hole mergers that, in turn, has important ramifications for data analysis applications with gravitational wave detectors and for modeling expensive merger simulations with reduced-order/surrogate models \cite{Field:2013cfa}.

Currently, we can answer the above question for spinning but {\it non-precessing} binary black hole coalescences, which involves only the three parameters $q$, $|\vec{\chi}_1|$, and $|\vec{\chi}_2|$, for which an effective-one-body (EOB) semi-analytical model of IMR is available \cite{BuonannoDamour:PRD59, Taracchini:2012ig}.
We first used our greedy algorithm to find the parameters for building a RB for the non-precessing inspiral PN waveforms using the $\varphi$-domain error in (\ref{eq:norm_phase}). 
We then generated a basis using the EOB non-precessing coalescence waveforms evaluated at those selected parameters. Lastly, we randomly generated more than $10^{d=3}$ EOB waveforms and computed the time-domain inertial frame mismatch from (\ref{eq:mismatch}). The results of this study are shown as the solid black curve in the right panel of Fig.~\ref{fig:errors}. For the first $20$ inspiral RB waveforms, the maximum mismatch of the EOB waveforms is about $3 \times 10^{-5}$ while for the first $50$ it is about $2 \times 10^{-7}$.

~\\
\noindent{\it Outlook}.--Based on traditional methods to sample the waveform space, which scale exponentially with dimension \cite{Owen:1995tm, Owen:1998dk, Harry:2009ea, Ajith:2012mn}, it has been perceived that an intractable number of numerical relativity simulations would be needed to represent the space of binary black holes for any given number of orbits. However, we have found evidence that a remarkably small number of numerical relativity binary black hole simulations may actually be needed, if {\it judiciously} chosen, to build a high accuracy reduced basis to represent the whole space of interest. 

Based on the non-precessing EOB results presented above, performing numerical simulations of binary black hole mergers for the first $50$-$90$ parameters selected by our greedy algorithm 
may be sufficient to represent the precession waveforms of any other coalescences in the parameter ranges of (\ref{eq:specs}). 
This constitutes less than one tenth of the number of randomly chosen simulations performed over the last few years  by the numerical relativity community
 \cite{Ajith:2012az, Hinder:2013oqa, Pekowsky:2013ska, Mroue:2013xna}.
In addition, this work suggests that an unexpectedly small number of low-mass inspiral waveforms may represent the frequency and parameter ranges of interest to gravitational wave detectors, which may also enable very compact reduced-order quadratures \cite{antil2012two,Canizares:2013ywa} of overlap integrals for fast parameter estimation studies. 
Finally, this work opens the door for building surrogate models \cite{Field:2013cfa} of precessing inspiral waveforms that can be useful for multiple query applications in place of solving a large number of parametrized ordinary differential equations.

~\\
\noindent{\it Acknowledgements}.--We thank Scott Field and Rory Smith for comments and discussions. We thank Michael Boyle for permission to use his Triton code for solving precessing PN waveforms, now replaced by his public open source code available from \cite{Triton}. We have also used the LAL (LSC Algorithm Library)  EOB code version 6.11.0.1 \cite{LAL_software}. This project was supported in part by the Fairchild Foundation, NSF grants PHY-1068881, CAREER PHY-0956189, and PHY-1005655 to Caltech, NASA grant NNX10AC69G, and NSF grants PHY-1208861, PHY-1316424 and PHY-1005632 to the University of Maryland. Computations were performed on the Zwicky cluster at Caltech, which is supported by the Sherman Fairchild Foundation and by NSF award PHY-0960291. Finally, we thank Ursula C.~T.~Gamma for continued inspiration.

\bibliographystyle{physrev}
\bibliography{references}

\end{document}